\documentclass[pra,twocolumn,showpacs,superscriptaddress,nofootinbib,a4paper]{revtex4}
\usepackage{amssymb}
\usepackage{amsbsy}
\usepackage{amsmath}
\usepackage[pdftex]{graphicx,color}
\usepackage{graphics}
\usepackage{graphicx}
\usepackage[colorlinks,linkcolor=blue]{hyperref}

\usepackage{mathptmx} 

\begin{document}

\title{Beyond  Adiabatic Elimination:  Systematic Expansions}

\author{I.~L. Egusquiza}
\affiliation{Department of Theoretical Physics and History of Science, UPV-EHU, 48080 Bilbao, Spain}

\begin{abstract}
We restate the adiabatic elimination approximation as the first term in a singular perturbation expansion. We use the invariant manifold formalism for singular perturbations in dynamical systems to identify systematic improvements on adiabatic elimination, connecting with well established quantum mechanical perturbation methods. We prove convergence of the expansions when energy scales are well separated. We state and solve the problem of  hermiticity of improved effective hamiltonians.
\end{abstract}




\maketitle
\section{Introduction}
Adiabatic elimination is a standard tool for the quantum optician \cite{walls2008quantum,Shore:2010uq,Yoo1985239,Steck:2013kx}. It is applied both to rate equations and amplitude equations. In its textbook presentation, the argument is that, because of wide separation of scales, the population or amplitude of a ``fast" state will not change appreciably from the point of view of a ``slow" state, if the initial state is mostly slow. The approximation then consists on setting the corresponding derivative to zero, solving for the fast variable, and introducing this solution in the rest of equations, thus obtaining an effective dynamics.

Given its widespread use, it natural that other presentations of the approximation appear. Of particular relevance is that achieved as the \emph{static} approximation in Laplace space, whereby an unknown energy term is discarded in a resolvent or Green function corresponding only to the fast sector. This also provides us with a suggestion for improving on the adiabatic approximation, as has been pointed out by Brion et al. \cite{1751-8121-40-5-011}, for instance.

In fact, the idea of separation of scales leading to approximations and effective dynamics is to be found in all areas of physics. In atomic and molecular physics one finds the Born--Oppenheimer approximation. In nuclear physics improvements on the shell model have led to effective interactions and operators to account for phenomena such as electric quadrupole moments. 

Yet again, effective lagrangians are a staple in nonperturbative analyses of field theories (see for instance chapter 12 of \cite{weinberg1996quantum}). One approach to understanding these effective lagrangians consists on carrying out the functional integral for the irrelevant degrees of freedom, thus producing a (generally non-local) effective lagrangian for the relevant degrees of freedom. Naturally enough, quite some skill is required to identify the proper degrees of freedom that need integrating out, and a number of techniques have been developed to this end. Additionally one normally needs to perform some approximations. One such combination of integrating out and approximation is again given by the \emph{static approximation}, which is the first term of an expansion of  a non-local propagator around a particular pole. 
This integration of irrelevant degrees of freedom and further approximation pertains to the use of field theory both for high energy and for condensed matter physics, differing in the symmetries and scales to be stressed.

The generality of this concept of separation of scales means that essentially the same technique has been often developed independently in different contexts. Although we most definitely do not attempt a  survey of each method in all areas, we do attempt to provide some pointers to the multiple sources of a particular perturbation procedure to which adiabatic elimination affiliates.

Because indeed adiabatic elimination in its textbook presentation can be understood as the first term in a systematic perturbative expansion, as we will show. Even more, we will show that it is an specific presentation of perturbation theory that is widespread in physics. The specifity, due to wide energy separations, will allow us furthermore a statement on convergence.

In this manner we will provide a response to the questions posed recently in, for example, \cite{1751-8121-40-5-011} and \cite{Paulisch:2012fk}: here the somewhat handwaving argument of standard adiabatic elimination is made precise, by identifying a small parameter that controls the precision of the approximation; adiabatic elimination forms part of a set of well-defined, consistent and convergent systematic approximations; the issues with the normalisation of the wave function and, in particular, with the computation of the population of the eliminated states are settled for each approximation; and, in appendix, we will detail the dependence on the interaction picture used to set up adiabatic elimination.

To do this we will first consider directly the well-known example of the \(\Lambda\) system, in which we will identify the relevant small parameter. We will proceed by recognising the system as a singular perturbation problem, which will be tackled with the invariant manifold formalism, the first term of the corresponding expansion being adiabatic elimination itself. We will recognise that in quantum mechanical systems this corresponds to linear embedding, and rewrite the method accordingly. 

In the next section we extend the findings for the \(\Lambda\) system to a general setting, and connect the results with the similarity formalism for perturbation theory that has been frequently rediscovered in different areas of physics. The third section proposes two systematic approximation schemes, and proves the existence of a solution. These schemes are illustrated in the next section by applying them again to the \(\Lambda\) system. 

Section  \ref{sec:hermapprox} is devoted to the issue of normalisation and hermiticity. It also connects again with the similarity formalism and effective diagonalisation. We end by proposing some conclusions and ideas for future work.

\section{The \(\Lambda\) System}
In order to compare results, we shall use the same notation as \cite{1751-8121-40-5-011}. Thus, the dynamical system of interest is
\begin{eqnarray}\label{eq:brionlambda}
i\dot\alpha&=& - \frac{ \delta}{2} \alpha+ \frac{ \Omega^*_a}{2} \gamma\nonumber\,,\\
i\dot \beta&=&  \frac{ \delta}{2} \beta+ \frac{ \Omega^*_b}{2} \gamma\,,\\
i\dot \gamma&=& \frac{ \Omega_a}{2} \alpha +\frac{ \Omega_b}{2} \beta+ \Delta \gamma\,.\nonumber
\end{eqnarray}
We are interested in the regime in which \(\Delta\gg \delta, \Omega_i\). The adiabatic elimination approximation consists of the argument that the \(\gamma\) component of the wave function will be approximately constant, and computed by solving for \(\gamma\) in the last equation after substituting \( \dot \gamma\to0 \). 

We shall first phrase the problem as a singular perturbation one, which will allow us to better understand the origin of adiabatic elimination. In order to do this, let us introduce the variable \( \tau= \delta t \), and define 
\(
\epsilon= \delta/ \Delta
\). The system is thus written as
\begin{eqnarray}
i\partial_\tau\alpha&=&- \frac{1}{2} \alpha+ \frac{ \Omega^*_a}{2 \delta} \gamma\nonumber\,,\\
i\partial_\tau\beta&=&\frac{1}{2} \beta+ \frac{ \Omega^*_b}{2 \delta} \gamma\,,\\
i \epsilon \partial_\tau\gamma&=& \gamma+ \frac{ \epsilon}{2}\left( \frac{ \Omega_a}{ \delta} \alpha + \frac{\Omega_b}{ \delta} \beta\right)\,.\nonumber 
\end{eqnarray}
This rewriting gives a direct justification for adiabatic elimination, by means of the \(\epsilon\) factor in front of \(\dot\gamma\). We can of course, go further.

For small \(\epsilon\) (and finite \( \Omega_i/ \delta \) for \( i= a,b \)), this is a singular perturbation problem: the problem, as stated, is a system with three differential equations, while for \(\epsilon=0\) it is a system with just two differential equations. Many methods have been developed for this kind of system \cite{kevorkian2011multiple,bender1999advanced}: averaging, multiple scales, dynamical renormalization group \cite{PhysRevLett.73.1311,PhysRevE.54.376} (for some applications of the dynamical renormalization group to quantum mechanics and quantum optics, see \cite{PhysRevA.57.1586} and \cite{PhysRevA.56.1548}). For our purposes we shall select the invariant manifold approach, which will be seen as a direct generalization of the standard adiabatic elimination process.

In the invariant manifold approach to singular perturbations, one determines among all possible manifolds invariant under the dynamics those that are perturbative. In our case, this is achieved by writing a submanifold of the three dimensional complex manifold in its explicit form \( \gamma=h( \alpha, \beta) \), and demanding that it be invariant under the dynamics. Denoting derivatives with respect to the variables \(\alpha\) and \(\beta\) by subindices, the invariance condition reads
\begin{multline}\label{eq:invariantlambda}
h+ \frac{ \epsilon}{2 \delta}\left( \Omega_a \alpha + \Omega_b \beta\right) =\\
\frac{ \epsilon}{2}\left[ \beta h_\beta- \alpha h_\alpha + \frac{h}{ \delta}\left( \Omega_a^* h_\alpha + \Omega_b^* h_\beta\right)\right]\,.
\end{multline}
Once the perturbation solution \( h \) has been determined, the effective evolution is 
\begin{eqnarray}
i\dot\alpha&=& - \frac{ \delta}{2} \alpha+ \frac{ \Omega^*_a}{2} h( \alpha, \beta)\nonumber\,,\\
i\dot \beta&=&  \frac{ \delta}{2} \beta+ \frac{ \Omega^*_b}{2} h( \alpha, \beta)\,.
\end{eqnarray}

The advantage of this method is that once a perturbative approximation has been determined for \( h \), the effective evolution will not present secular terms. The disadvantage is that the effective evolution has to be recomputed for different approximations.

In the case at hand, the perturbative solution of eqn. (\ref{eq:invariantlambda}) has the form \( h=\sum_{k=1}^\infty \epsilon^k h_k \). Thus
\begin{eqnarray}
h_1( \alpha, \beta)&=& - \frac{1}{2 \delta}\left( \Omega_a \alpha+ \Omega_b \beta\right)\nonumber\,,\\
h_2( \alpha, \beta)&=& \frac{1}{4 \delta}\left( \Omega_a \alpha- \Omega_b \beta\right)\nonumber\,,
\end{eqnarray}
and, for \( k\geq2 \),
\[
h_{k+1}= \frac{1}{2}\left( \beta\partial_\beta- \alpha\partial_\alpha\right)h_k + \frac{1}{2 \delta}\sum_{l=1}^{k-1} h_{k-l}\left( \Omega_a^*\partial_\alpha+ \Omega_b^*\partial_\beta\right) h_l\,.
\]
The first term in the expansion, \( h_1 \), corresponds exactly with that is termed adiabatic elimination, thus making clear in which way the adiabatic elimination approximation is a first term in a (singular) perturbative expansion, as desired from the outset.

The invariant manifold approach is specially adequate for nonlinear dynamical systems, and it can be simplified in the case at hand:
even though the embedding equation (\ref{eq:invariantlambda}) is nonlinear, the perturbative solution is linear (one way of seeing this is that the differential system is the characteristic system for the embedding partial differential equation). Therefore, the effective evolution on the invariant manifold is also linear; it is, however, not unitary. This comes about because the conserved quantity is actually the full norm \( | \alpha|^2+ | \beta|^2+ | \gamma|^2 \), and not just  \( | \alpha|^2+ | \beta|^2\).

Since the embedding is linear, we can rephrase it as 
\[ \gamma= \left(b_1,b_2\right) \begin{pmatrix}
\alpha\\ \beta
\end{pmatrix}= B \begin{pmatrix}
\alpha\\ \beta
\end{pmatrix}\,.\]
\( B \) is an operator from the two dimensional space spanned by \( \left(1,0,0\right)^{\dag} \) and  \( \left(0,1,0\right)^{\dag} \) to the one dimensional space spanned by  \( \left(0,0,1\right)^{\dag} \). The invariance condition reads now 
\begin{equation}\label{eq:invariblambda}
-\frac{ \delta}{2 \Delta} B \sigma_3+ \frac{1}{ \Delta} B \Omega^{\dag} B= B + \frac{1}{ \Delta} \Omega\,,
\end{equation}
where
\[ \Omega = \frac{1}{2}\left( \Omega_a, \Omega_b\right)\,.\]
Given any \( B \) solution of the invariance condition (\ref{eq:invariblambda}), a non hermitian effective Hamiltonian is computed as 
\[h_{ \mathrm{eff}}= - \frac{ \delta}{2} \sigma_3+ \Omega^{\dag} B\,.\]
\section{General Formal Theory}
\subsection{Direct approach}
The presentation above inspires a more general, admittedly formal, theory. Consider a Hilbert space \( \mathcal{H} \) on which a Hamiltonian acts, with clearly separate scales present (in a sense to be made clear later). Assume therefore that there is a subspace \( \mathcal{H}_P \) corresponding to the \emph{slow} variables (also called the model or valence space in  nuclear physics, or the active space in atomic and molecular physics). The complementary space, \( \mathcal{H}_Q= \mathcal{H}_P^\perp \), with \( Q=1-P \), is used to describe the \emph{fast} variables. The hamiltonian can be written in block diagonal form,
\begin{equation}
H= \begin{pmatrix}
PHP&PHQ\\ QHP& QHQ
\end{pmatrix}= \hbar \begin{pmatrix}
\omega &\Omega^{\dag}\\
\Omega& \Delta
\end{pmatrix}\,,
\end{equation}
which corresponds to the direct sum structure \( \mathcal{H}= \mathcal{H}_P \oplus \mathcal{H}_Q \) as
\[ \mathcal{H} \ni\psi= \begin{pmatrix}
P\psi\\ Q\psi
\end{pmatrix}= \begin{pmatrix}
\alpha \\ \gamma
\end{pmatrix}\,.\]
The embedding condition means the identification of manifolds invariant under the flow determined by the Hamiltonian, 
\[i \partial_t \begin{pmatrix}
\alpha\\ \gamma
\end{pmatrix}= \begin{pmatrix}
\omega &\Omega^{\dag}\\
\Omega& \Delta
\end{pmatrix}\begin{pmatrix}
\alpha\\ \gamma
\end{pmatrix}\,,\]
with the specification that the invariant manifold is isomorphic (unitarily equivalent) to \( \mathcal{H}_P \) (for finite dimensionality, of the same dimensionality). This can be achieved in an explicit form by means of operators \( B: \mathcal{H}_P\to \mathcal{H}_Q \), which obey the equation
\begin{equation}\label{eq:bembedding}
\Omega+ \Delta B= B \omega + B \Omega^{\dag} B\,.
\end{equation}
To prove the need for this condition, substitute \( \gamma = B \alpha \) in the evolution equation, and ask for consistency.

The effective evolution for \(\alpha\) is readily seen to be \( i\dot\alpha=\left( \omega+ \Omega^{\dag}B\right) \alpha \), thus suggesting the effective hamiltonian 
 \( h_{ \mathrm{eff}}= \omega+ \Omega^{\dag}B \).

This is in fact the equation for what has been called in the literature the reduced Bloch wave operator or the \emph{model} or \emph{wave} operator from nuclear physics \cite{PhysRev.97.1366,Ellis:1977kx} (see also section 2 of \cite{0305-4470-36-20-201} and appendix \ref{app:blochwave}). We have obtained the equation directly from reduction by partitioning of Schr\"odinger's time dependent equation, while the reduced wave operator is normally obtained from the eigenvalue equation applying the same partitioning, following the projector operator formalism of Feshbach \cite{Feshbach1958357,Feshbach1962287} (notice the reprinting of the latter as \cite{Feshbach2000519}). The advantage of our approach will come from it  making visible the singular perturbation aspect of the equations in the adiabatic elimination case.

An important aspect here is that  in principle the embedding algebraic equation (\ref{eq:bembedding}) has solutions that give the \emph{full} spectrum of the total hamiltonian. For instance, if the slow and fast variables were decoupled, with no degeneracy, the trivial solution \( B=0 \) would be unique. As a consequence of this equation involving all eigenspaces of the full hamiltonian, it can lead in some circumstances to problems in its perturbative solution, related to the existence of so-called intruder states, for instance. 

\subsection{Similarity formalism}\label{ssec:similarity}

Alternatively, it is also related to the similarity transformation formalism, as follows. As is known, similarity transformations are isospectral. We thus look for invertible operators \( X \) such that 
\begin{equation}\label{eq:decoupling}
 Q X^{-1} H X P=0\,,
\end{equation}
and some additional conditions are also fulfilled.
Given such an operator \( X \), the effective hamiltonian in the slow or model space would be given by 
\[H_X^P= P X^{-1} H X P\,.\]
Notice that \( H_X^P \) need not be hermitian; it is indeed hermitian if \( X \) is unitary. A bonus of this approach is that effective operators \( A_{ \mathrm{eff}} \) acting on \( \mathcal{H}_P \) are constructed for each \( A \) acting on the full Hilbert space \( \mathcal{H} \) as 
\[ A_{ \mathrm{eff}}= P X^{-1} A X P\,.\]

This similarity transformation construction, without further restriction, would not be very informative (for instance, it need not reflect an existing hierarchy of scales), and thus a number of additional conditions have been developed in the literature \cite{Suzuki01071982,Suzuki01081983} to best reflect the underlying physics. When Van Vleck first introduced this formalism in 1929 (section 4 of \cite{Vleck:1929fk}) he demanded unitarity of \( X \). This approach was reintroduced in relativistic quantum mechanics by Foldy and Wouthuysen 
\cite{PhysRev.78.29}, and in condensed matter physics some time later by Schrieffer and Wolff \cite{PhysRev.149.491} (for a review, see \cite{Bravyi20112793}).

If unitarity is not demanded, an immediate proposal is 
\[X_B= \exp\left(B\right)\,.\]
Here we have a slight overload of notation: we have used \( B \) as an operator acting on \( \mathcal{H} \), taking into account that \( B=QBP=QB=BP \), which in turn implies
\[X_B= 1+B\]
(since \( B^2=0 \))
and 
\[X_B^{-1}= 1-B\,.\]
Inserting these in the decoupling condition (\ref{eq:decoupling}), we have 
\begin{multline}
Q\left(1-B\right)H\left(1+B\right)P=
\left(Q-B\right)H\left(P+B\right)=\\
QHP-BPHP+QHQB-BPHQB=\\
\hbar\left[ \Omega - B \omega + \Delta B - B \Omega^{\dag} B\right]\,,
\end{multline}
thus proving that with this choice of \( X \) the decoupling condition is precisely the embedding equation.

With this proposal, for a solution of the embedding equation we have an effective hamiltonian \( h_{ \mathrm{eff}}= \omega+ \Omega^{\dag}B \), which is generically not hermitian. This effective hamiltonian has been called the Bloch hamiltonian in the literature \cite{0305-4470-36-20-201}.

\section{Formal Adiabatic Elimination Expansion and Convergence}
As stated above, the embedding equation is by itself very general, and can have multiple solutions. We want to study the specific case in which, physically, there is a wide separation of scales. For the \(\Lambda\) system this pertained to the condition \( \Delta\gg \delta, \Omega_a, \Omega_b \). In general, we will achieve separation of scales if our choice of subspaces (equivalently of the \( P \) projector) is such that a) \(\Delta\) is invertible as an operator on \( \mathcal{H}_Q \), b) 
\[ \epsilon=\| \Delta^{-1}\| \| \omega\|\ll1\,,\]
and c)
\[ \epsilon'= \| \Delta^{-1}\| \| \Omega\|\ll1\,,\]
for  operator norms.

Define a new time variable \( \tau= \| \omega\| t \). The equations of motion are now 
\begin{eqnarray}
i \partial_\tau \alpha&=& \frac{ \omega}{\| \omega\|} \alpha+ \frac{ \Omega^{\dag}}{\| \omega\|} \gamma\,,\nonumber\\
i \epsilon \partial_\tau \gamma&=& \epsilon \frac{ \Omega}{\| \omega\|} \alpha+ \| \Delta^{-1}\| \Delta \gamma\,,
\end{eqnarray}
in which the singular perturbation character is clearly visible. This suggests treating the embedding equation as determining either a perturbative expansion or a recurrence relation.

If indeed \(\Delta\) is invertible as an operator on \( \mathcal{H}_Q \), the embedding condition (\ref{eq:bembedding}) can be rewritten as
\[B=T(B)\,,\]
where \( T \) maps \( \mathcal{B}\left( \mathcal{H}_P,\mathcal{H}_Q \right)\) (bounded operators from \( \mathcal{H}_P \) to  \( \mathcal{H}_Q \)) to itself nonlinearly, as follows 
\begin{equation}
T(A)= - \Delta^{-1} \Omega + \Delta^{-1}A \omega + \Delta^{-1} A \Omega^{\dag} A\,.
\end{equation}
For any norm on \( \mathcal{B}\left( \mathcal{H}_P,\mathcal{H}_Q \right)\) we have the bound
\[
\|T(A)\|\leq \|\Delta^{-1} \Omega\| + \|\Delta^{-1}A \omega\| + \|\Delta^{-1} A \Omega^{\dag} A\|\,.
\]
For the operator norm, we also have 
\begin{multline}
\|T(A)\|\leq \|\Delta^{-1}\|\,\| \Omega\| + \|\Delta^{-1}\|\,\|  \omega\|\,\| A\| + \|\Delta^{-1} \|\,\| \Omega\|\,\| A\|^2\\
\leq \epsilon'\left(1+  \|A\|^2\right) + \epsilon \|A\|\,.
\end{multline}
Now, by plotting the functions \( x \) and \( g(x)= \epsilon'(1+x^2)+ \epsilon x \) for \( \epsilon, \epsilon'\geq0 \) and \( \epsilon'\leq (1- \epsilon)/2 \), one readily sees that if \( A \) is in the ball centered  at the null operator with radius \( r( \epsilon, \epsilon')= (1- \epsilon)/(2 \epsilon')+ \sqrt{ (1- \epsilon)^2/(2 \epsilon')^2-1} \) then \( T(A) \) also belongs to that ball. Thus, by Schauder's fixed point theorem, \( T \) has a fixed point in the ball. Let us now construct the recurrence
\begin{eqnarray}\label{eq:brecurrence}
B^{(0)}&=& - \Delta^{-1} \Omega\,,\\
B^{(k+1)}&=& T\left[B^{(k)}\right]\nonumber\,.
\end{eqnarray}
Since \( \|B^{(0)}\|\leq \epsilon' \), it will belong to the ball if \( \epsilon'\leq r( \epsilon, \epsilon') \) (which is the generic situation), and the recurrence will converge (we have actually not excluded the possibility that there exist a fixed cycle --- see however Appendix \ref{app:nocycles}).

Furthermore, writing \( \epsilon'= a \epsilon \), where \( a= \| \Omega\|/\| \omega\| \) is a fixed constant, we can write the embedding equation as perturbative, where \( \Delta^{-1} \) has perturbative weight \(\epsilon\): expand
\begin{equation}\label{eq:adiaelimexp}
B= \sum_{k=1}^\infty B_{(k)}
\end{equation}
and insert in the embedding equation to obtain
\begin{eqnarray}\label{eq:adiaelimexprecurr}
B_{(1)}&=&- \Delta^{-1} \Omega\,,\nonumber\\
B_{(2)}&=& - \Delta^{-2} \Omega \omega\,,\\
B_{(k+1)}&=& \Delta^{-1} B_{(k)} \omega + \Delta^{-1} \sum_{l=1}^{k-1}B_{(k-l)} \Omega^{\dag} B_{(l)}\,,\nonumber
\end{eqnarray}
where in the last line \( k\geq2 \). Notice that, formally,
\[B^{(k)}-\sum_{l=1}^{k+1}B_{(l)}=O\left( \Delta^{-(k+2)}\right)\,.\]

We shall give the name \emph{perturbative adiabatic elimination expansion} to the expansion (\ref{eq:adiaelimexp}) with recurrence (\ref{eq:adiaelimexprecurr}).
Given this expansion to some order \( k \), we shall have the effective hamiltonian to the same order by substituting the approximation for \( B \) in  \( h_{ \mathrm{eff}}= \omega+ \Omega^{\dag}B \).

Notice that we have the expansion in a purely algebraic manner; there is no need to impose consistency through the single pole approximation for the resolvent, as in \cite{1751-8121-40-5-011}, or, equivalently, in the Taylor expansion proposed in \cite{Paulisch:2012fk}.

One must bear in mind that this is a particular perturbation expansion, particularly adapted to the problem at hand, that differs from other expansions present in the literature. See appendix \ref{app:seconperturb} for another perturbative expansion, which is adequate even if  our expansion parameter is not small.

\section{Example: the \(\Lambda\) system}
We can now identify the constructions of the previous section in the \(\Lambda\) system previously introduced; as one can see
\[ \omega= - \frac{ \delta}{2} \sigma_3\,,\quad \Omega= \frac{1}{2}\left( \Omega_a, \Omega_b\right)\,,\]
and \(\Delta\) is simply a number.

To fourth order we have
\begin{eqnarray}
B_{(1)}&=&- \frac{1}{ \Delta} \Omega\,,\nonumber\\
B_{(2)}&=& \frac{ \delta}{2 \Delta^2} \Omega \sigma_3\,,\\
B_{(3)}&=& \frac{1}{ \Delta^3}\left( \frac{ \delta}{2}+ \Omega \Omega^{\dag}\right) \Omega\,,\nonumber\\
B_{(4)}&=& - \frac{ \delta}{2 \Delta^4}\left( \Omega \sigma_3 \Omega^{\dag}\right) \Omega - \frac{ \delta}{2 \Delta^4}\left( 2 \Omega \Omega^{\dag}+ \frac{ \delta}{2}\right) \Omega \sigma_3\,.\nonumber
\end{eqnarray}
Notice that at any order \( B_{(k)} \) will generically have a term proportional to the operator \(\Omega\) and another proportional to \(\Omega \sigma_3\), with no other possibility.

To the same order we obtain
\begin{eqnarray}
h_{ \mathrm{eff}} &=& - \frac{ \delta}{2} \sigma_3 \nonumber\\
&&- \frac{1}{ \Delta} \left( 1- \frac{ \delta}{2 \Delta^2}- \frac{ \Omega \Omega^{\dag}}{ \Delta^2}+ \frac{ \delta}{2 \Delta^3} \Omega \sigma_3 \Omega^{\dag}\right) \Omega^{\dag} \Omega\nonumber\\
&& + \frac{ \delta}{2 \Delta^2} \left(1- \frac{ \delta}{2 \Delta^2}+ \frac{ \Omega \Omega^{\dag}}{ \Delta^2}\right) \Omega^{\dag} \Omega \sigma_3+ \cdots\nonumber
\end{eqnarray}
This expression has been ordered in terms of the operators \( \sigma_3 \), \(  \Omega^{\dag} \Omega \) and \(  \Omega^{\dag} \Omega \sigma_3 \), with numerical coefficients in front; again, these are the only possibilities. The non hermitian terms are due to \(  \Omega^{\dag} \Omega \sigma_3 \). 
Notice that some terms with physical significance are spread in more than one coefficient; for instance, the Stark shift has contributions from both \( \mathrm{Tr}\left[ \Omega^{\dag} \Omega\right] \) and  \( \mathrm{Tr}\left[ \Omega^{\dag} \Omega \sigma_3\right] \).

\begin{figure}[htbp]
\begin{center}
\includegraphics[width=2in]{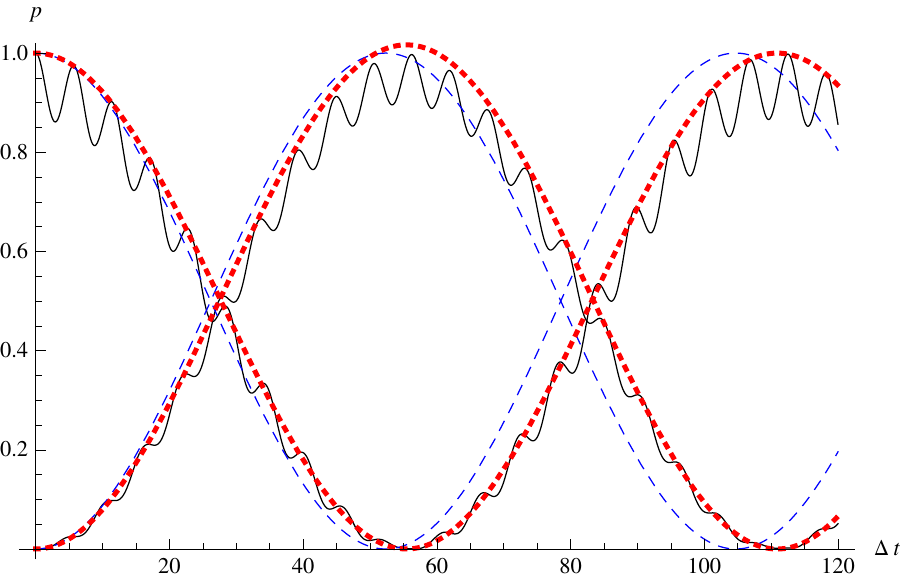}

\caption{Evolution of the population of the ground and excited states, with initial state \( (1,0,0)^T \), under a) the exact hamiltonian (continuous black line), b) zeroth order effective hamiltonian (dashed blue line) and c) fourth iteration of \( T \) (dotted red line). The parameters are \( \delta= -0.0175 \Delta\), \( \Omega_a=0.4 \Delta \), \( \Omega_b=0.3 \Delta \), for direct comparison with \cite{Paulisch:2012fk}.}
\label{fig:lambdaex}
\end{center}
\end{figure}

Numerically it will be faster to use the recurrence relation. In fig. \ref{fig:lambdaex} we depict the evolution of the populations in the slow manifold for exact evolution, effective hamiltonian with \( B^{(0)} \), and effective hamiltonian constructed with \( B^{(4)} \). At the scale of the figure, the graphs constructed with \( B^{(2)} \) and \( B^{(3)} \) would be indistinguishable from the fourth iteration for those values of the parameters. Notice that the hamiltonian \( h_{ \mathrm{eff}}^{(0)}= \omega+ \Omega^{\dag} B^{(0)} \) is the usual adiabatically eliminated hamiltonian, and that  \(  h_{ \mathrm{eff}}^{(4)}\) is much more successful in tracking the envelope of the evolution. For the numerical values used, the non conservation of probability due to the nonhermiticity of \(  h_{ \mathrm{eff}}^{(4)}\) is of order \( 5\% \).

\section{Hermitian approximations: a general theory}\label{sec:hermapprox}

\subsection{Exact effective hermitian hamiltonian}

As stated above, the nonhermiticity of Bloch's hamiltonian \( h_{ \mathrm{eff}}= \omega+ \Omega^{\dag} B \) is due to the part of the norm carried by \( \gamma=B \alpha\in \mathcal{H}_Q \). The norm conserved under the evolution with the full hamiltonian in \( \mathcal{H} \) is simply
\begin{multline}
\left\langle \alpha, \alpha\right\rangle_{ \mathcal{H}_P}+\left\langle \gamma, \gamma\right\rangle_{ \mathcal{H}_Q}=\\
\left\langle \alpha, \alpha\right\rangle_{ \mathcal{H}_P}+\left\langle B \alpha, B \alpha\right\rangle_{ \mathcal{H}_Q}=\\
\left\langle \alpha, \alpha\right\rangle_{ \mathcal{H}_P}+\left\langle \alpha, B^{\dag}B \alpha\right\rangle_{ \mathcal{H}_P}=\\
\left\langle \alpha,\left(1+ B^{\dag}B\right) \alpha\right\rangle_{ \mathcal{H}_P}\,.
\end{multline}
By construction, \( 1+ B^{\dag}B \) is positive, and one can define its square root
\[S_B= \sqrt{1+ B^{\dag}B }\,.\]
Then for any constant unitary \( V \) from \( \mathcal{H}_P \) to itself we have that the norm of 
\[ \tilde{\alpha}_V= V S_B \alpha\]
is conserved if \(\alpha\) evolves under the Bloch hamiltonian \( h_{ \mathrm{eff}} \). I.e.
\[i\partial_t\tilde{ \alpha}_V=V S_B i\partial_t \alpha= V S_B h_{ \mathrm{eff}} S_B^{-1}V^{-1} \tilde{ \alpha}_V\,,\]
from which one defines the family of hamiltonians
\[
h_V= V S_B h_{ \mathrm{eff}} S_B^{-1}V^{-1} \,,
\]
which are indeed hermitian. This last statement can be proven in several ways. Firstly, by construction: the flow under \( h_V \) preserves the norm and the inner products. It is a continuous flow, so by Wigner's theorem it is a unitary flow. The hamiltonians \( h_V \) are the constant generators of the unitary flow, and thus hermitian. In finite dimensions we can also assert self-adjointness.

A second proof is explicit. Observe that
\[ S_B h_{ \mathrm{eff}} S_B^{-1}= S_B^{-1}\left(1+B^{\dag}B\right) h_{ \mathrm{eff}} S_B^{-1}\,.\]
Since \( S_B \) is hermitian, we will have proven hermiticity of \( h_V \) if we prove hermiticity of \( \left(1+B^{\dag}B\right) h_{ \mathrm{eff}}  \), that is, of
\[\left(1+B^{\dag}B\right)\left( \omega+ \Omega^{\dag} B\right)\,.\]
And if \( B \) is a solution of eqn. (\ref{eq:bembedding}) we indeed have that
\begin{eqnarray}\label{eq:computehermit}
\left(1+B^{\dag}B\right)\left( \omega+ \Omega^{\dag} B\right)&=& \omega + \Omega^{\dag} B+ B^{\dag}\left(B \omega+ B \Omega^{\dag}B\right)\nonumber\\
&=& \omega + \Omega^{\dag} B+ B^{\dag}\left( \Delta B+ \Omega\right)\\
&=& \omega + \Omega^{\dag} B+ B^{\dag} \Omega + B^{\dag} \Delta B\nonumber\,,
\end{eqnarray}
which, given the hermiticity of \(\omega\) and \(\Delta\), is explicitly hermitian. Notice that we have used eqn.  (\ref{eq:bembedding}) in the second step.

\subsection{Approximate effective hermitian hamiltonians}
This proof of hermiticity relies on \( B \) being an exact solution of (\ref{eq:bembedding}); if \( B_{\mathrm{a}} \)  were an approximation to the solution to a given order, then 
\begin{equation}
\sqrt{1+ B_{\mathrm{a}}^{\dag}B_{\mathrm{a}} } \left( \omega + \Omega^{\dag} B_{\mathrm{a}}\right) \frac{1}{\sqrt{1+ B_{\mathrm{a}}^{\dag}B_{\mathrm{a}} } }\nonumber
\end{equation}
will be approximately hermitian, to the same order. However, the proof itself suggests that
\begin{equation}\label{eq:approxhermitian}
h_V\left[B_{ \mathrm{a}}\right]= \frac{1}{\sqrt{1+ B_{\mathrm{a}}^{\dag}B_{\mathrm{a}}}} \left( \omega + \Omega^{\dag} B_{\mathrm{a}}+ B_{\mathrm{a}}^{\dag} \Omega + B_{\mathrm{a}}^{\dag} \Delta B_{\mathrm{a}} \right)\frac{1}{\sqrt{1+ B_{\mathrm{a}}^{\dag}B_{\mathrm{a}}}}
\end{equation}
is a always a hermitian approximation.

In particular, by using the successive approximations \( B^{(k)} \) introduced in (\ref{eq:brecurrence}) one can define  successive hermitian approximations of the full hermitian effective hamiltonian as
\[h_V^{(k)}=h_V\left[B^{(k)}\right]\,.\]

\begin{figure}[htbp]
\begin{center}
\includegraphics[width=2in]{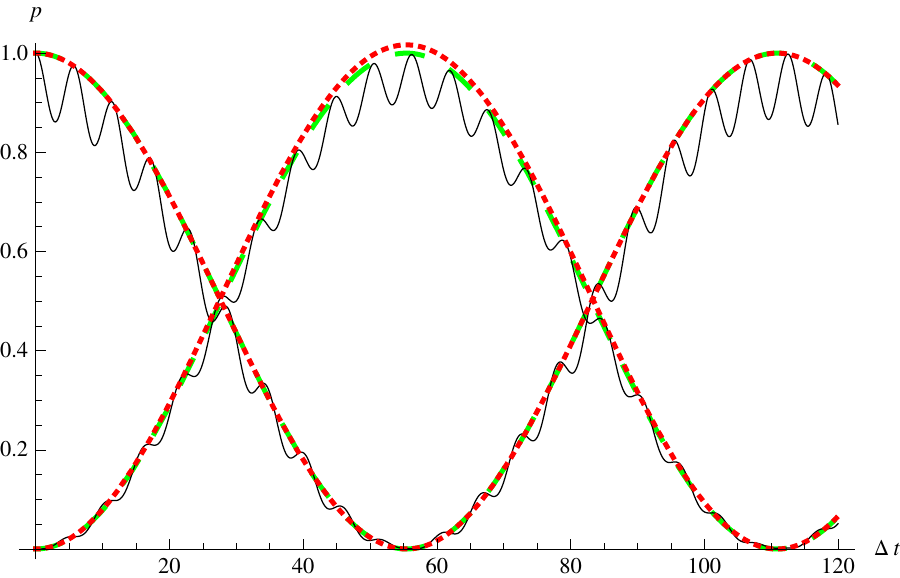}

\caption{Evolution of the population of the ground and excited states, with initial state \( (1,0,0)^T \), under a) the exact hamiltonian (continuous black line), b) \( h_{ \mathrm{eff}}^{(4)} \) (dotted red line) and c) \( h_V^{(10)} \) with \( V=1 \) (dashed green line). The parameters are \( \delta= -0.0175 \Delta\), \( \Omega_a=0.4 \Delta \), \( \Omega_b=0.3 \Delta \), as before.}
\label{fig:lambdaherm}
\end{center}
\end{figure}

As an example, in figure \ref{fig:lambdaherm} we depict also the evolution with the hermitian hamiltonian, again for the \(\Lambda\) system. Observe that the central maximum is at one, and that norm is conserved.
\subsection{Effective triangularization/diagonalisation}
In the previous subsections we have emphasised the effective hamiltonian, be it exact or approximate, in the slow sector. However, the construction provides us also with information about the fast sector. 

Look back to subsection \ref{ssec:similarity}. There we defined
\[H_X^P= P X^{-1} H X P\]
as the effective hamiltonian in the slow space. We could similarly regard
\[H_X^Q= Q X^{-1} H X Q\]
as the effective hamiltonian in the fast space.  This is correct in that it provides us with the eigenvalues corresponding to the fast space: since we impose the condition 
\(Q X^{-1} H X P=0\), the similarity transformed \( X^{-1} H X \) is block upper triangular, and, since the eigenvalues of a block upper triangular matrix are the eigenvalues of its diagonal blocks, the fast eigenvalues are indeed those of \( H_X^Q \). 

Using \( X_B \), with \( B \) an exact solution of the embedding condition (\ref{eq:bembedding}), we have
\[X_B^{-1} \begin{pmatrix}
\omega& \Omega^{\dag}\\ \Omega & \Delta
\end{pmatrix} X_B= \begin{pmatrix}
\omega+ \Omega^{\dag}B & \Omega^{\dag}\\0& \Delta- B \Omega^{\dag}
\end{pmatrix}\,,\]
so that \( \Delta- B \Omega^{\dag} \) is the effective hamiltonian in the fast space.

We now desire to have a hermitian effective hamiltonian in the fast sector. In order to do that, let us first notice that,
similarly to the computation in (\ref{eq:computehermit}), one can prove that
\[\left(1+B B^{\dag}\right)\left( \Delta- \Omega B^{\dag}\right)= \Delta- \Omega B^{\dag}- B \Omega^{\dag}+ B \omega B^{\dag}\,,\]
which is explicitly hermitian, as long as \( B \) is a solution of eqn. (\ref{eq:bembedding}). Define 
\[\tilde{S}_B= \sqrt{1+ B B^{\dag}}\]
acting on \( \mathcal{H}_Q \) (for later computations it will prove useful to bear in mind that \( \tilde{S}_B^2 B= B S_B^2 \), and thus \( \tilde{S}_B B S_B^{-1}= \tilde{S}_B^{-1} B S_B\)). We have proven that \( H \) is similar to a block upper triangular matrix with hermitian diagonal blocks and bounded off-diagonal block. Explicitly: the computations hitherto signify 
\begin{multline}\label{eq:vssimilarity}
\begin{pmatrix}
V_P S_B&0\\-V_Q \tilde{S}_B B& V_Q \tilde{S}_B
\end{pmatrix} \begin{pmatrix}
\omega& \Omega^{\dag}\\ \Omega& \Delta
\end{pmatrix}
\begin{pmatrix}
S_B^{-1}V_P^{\dag}&0\\B S_B^{-1}V_P^{\dag}& \tilde{S}_B^{-1}V_Q^{\dag}
\end{pmatrix}
=\\
\begin{pmatrix}
V_P h_c^\alpha V_P^{\dag} & V_P S_B \Omega^{\dag}\tilde{S}_B^{-1}V_Q^{\dag}\\
0& V_Q h_c^\gamma V_Q^{\dag} 
\end{pmatrix}\,,
\end{multline}
where
\begin{eqnarray}
h_c^\alpha&=& \frac{1}{\sqrt{1+ B^{\dag}B}}\left[\omega + \Omega^{\dag} B+ B^{\dag} \Omega + B^{\dag} \Delta B\right]  \frac{1}{\sqrt{1+ B^{\dag}B}}\,,\nonumber\\
h_c^\gamma&=& \frac{1}{\sqrt{1+ BB^{\dag}}}\left[\Delta- \Omega B^{\dag}- B \Omega^{\dag}+ B \omega B^{\dag}\right]  \frac{1}{\sqrt{1+ BB^{\dag}}}\,.\nonumber\
\end{eqnarray}
Notice that 
\[\begin{pmatrix}
S_B^{-1}V_P^{\dag}&0\\B S_B^{-1}V_P^{\dag}& \tilde{S}_B^{-1}V_Q^{\dag}
\end{pmatrix}\]
is the inverse of 
\[
\begin{pmatrix}
V_P S_B&0\\-V_Q \tilde{S}_B B& V_Q \tilde{S}_B
\end{pmatrix} 
\]
with \( V_P \) and \( V_Q \) unitaries acting on \( \mathcal{H}_P \) and  \( \mathcal{H}_Q \) respectively, so that the left hand side of (\ref{eq:vssimilarity}) is indeed a similarity transformation.

One of the advantages of this approach to the problem is that we have  approximations for \( h_c^\alpha \) and \( h_c^\gamma \)  which are explicitly hermitian for each approximation of \( B \). So we have a block upper triangular matrix that is an approximation to the exact one for each approximation of \( B \) solution of the embedding equation (\ref{eq:bembedding}), with hermitian diagonal blocks and bounded off-diagonal block.

 In finite dimensions, Roth's theorem \cite{MR0047598} gives us a condition for this block upper triangular matrix to be in turn similar to the block diagonal matrix with the same diagonal blocks. For infinite dimensions, if we can guarantee self-adjointness and not just hermiticity of the diagonal blocks, we can make use of the analogous result by Rosenblum \cite{MR0233214}. 
 
 The simple result, from our point of view, is that if for a given \( B \) the spectra of \( h_c^\alpha(B) \) and  \( h_c^\gamma(B) \) are disjoint, then a similarity transformation exists that transforms the block upper triangular matrix into the corresponding block diagonal one. This similarity transform is given by \( Y: \mathcal{H}_Q \to \mathcal{H}_P\), solution of  the Sylvester equation
 \[h_c^\alpha Y-Y h_c^\gamma= S_B \Omega^{\dag} \tilde{S}_B^{-1}\]
through
\[ \begin{pmatrix}
1&Y\\0&1
\end{pmatrix}
\begin{pmatrix}
h_c^\alpha& S_B \Omega^{\dag} \tilde{S}_B^{-1}\\0& h_c^\gamma
\end{pmatrix}
\begin{pmatrix}
1&-Y\\0&1
\end{pmatrix}=\begin{pmatrix}
h_c^\alpha& 0\\0& h_c^\gamma
\end{pmatrix}\,.\]

The crucial point for us at this juncture is to note that the existence of the diagonalising transformation (that must necessarily be unitary and not just by similarity) is guaranteed if the spectra of \( h_c^\alpha(B) \) and  \( h_c^\gamma(B) \) are disjoint; this is intuitively the case if indeed \(\omega\) and \(\Omega\) are much smaller than \(\Delta\) in the sense above.

Actually, there is a much faster explicit route: notice that
\begin{multline}
\begin{pmatrix}
1& B^{\dag}\\ -B&1
\end{pmatrix}
\begin{pmatrix}
\omega& \Omega^{\dag}\\ \Omega& \Delta
\end{pmatrix}
\begin{pmatrix}
1& -B^{\dag}\\ B&1
\end{pmatrix}=\\
\begin{pmatrix}
\omega+ B^{\dag} \Omega+ \Omega^{\dag}B+ B^{\dag} \Delta B&0\\0& \Delta- \Omega B^{\dag}-B \Omega^{\dag}+ B \omega B^{\dag}
\end{pmatrix}\,,\end{multline}
if \( B \) is a solution of eqn. (\ref{eq:bembedding}). Define 
\begin{eqnarray}\label{eq:finalx}
X= \begin{pmatrix}
1& -B^{\dag}\\ B&1
\end{pmatrix} \begin{pmatrix}
S_B^{-1}&0\\0& \tilde{S}_B^{-1}
\end{pmatrix}\,.
\end{eqnarray}
Then 
\[X^{-1}= \begin{pmatrix}
S_B^{-1}&0\\0& \tilde{S}_B^{-1}
\end{pmatrix}\begin{pmatrix}
1& B^{\dag}\\ -B&1
\end{pmatrix}\,,\]
and we have proved that, for \( B \) solution of eqn. (\ref{eq:bembedding}), 
\[X^{-1} \begin{pmatrix}
\omega& \Omega^{\dag}\\ \Omega& \Delta
\end{pmatrix} X= \begin{pmatrix}
h_c^\alpha& 0\\0& h_c^\gamma
\end{pmatrix}\,.\]

There is a subtlety to be pointed out for the sake of the unwary: the fact that \( X \) constructed as in definition (\ref{eq:finalx}) gives the same diagonalization as 
\[X_Y= 
\begin{pmatrix}
1&0\\B&1
\end{pmatrix}
\begin{pmatrix}
S_B^{-1}&0\\0& \tilde{S}_B^{-1}
\end{pmatrix}
\begin{pmatrix}
1&-Y\\0&1
\end{pmatrix}\]
does not mean that they are the same. Rather, it means that \( X X_Y^{-1} \) commutes with \( H \).

The explicit expression for \( X \) given here was first presented in \cite{Suzuki01071982}. The alternative expression 
\[X=\exp\left[ \mathrm{arctanh}\left(B-B^{\dag}\right)\right]\] has been computed independently by several authors (additional to \cite{Suzuki01071982,Suzuki01081983}, see also \cite{Bravyi20112793}, more directly adapted to the presentation of Schrieffer and Wolff \cite{PhysRev.149.491}).

In the context of quantum optics, the matrix \( X \) provides us with the dressing of the subspaces:
\[XPX^{-1}=P_X\]
is the projector onto the low lying sector. In other words, \( X \) intertwines the projector onto the unperturbed low lying states and the projector onto the fully non-perturbative low lying states, \(XP=P_X X \).

Notice that given any \( B \) (solution of eqn. (\ref{eq:bembedding}) or not), \( X \) is unitary, and it fully determines a projector \( P_X \). This means that if we have an approximate solution of eqn. (\ref{eq:bembedding}) we have an approximation to the projector onto the space of low lying states. Similarly, we will also have an approximation for the effective hamiltonian, \( h_c^\alpha \), hermitian by construction. 

In the case of interest, where two subspaces are tentatively identified as corresponding to widely separated time scales, this approximation can be given by truncating at a given order the perturbative adiabatic elimination expansion, or, alternatively, by the recurrence relation.
\section{Conclusions}
We have identified the adiabatic elimination approximation as the first member of sets of systematic approximations, built algebraically in a well defined way. In fact, we have made explicit the connection between adiabatic elimination and well established perturbation methods. As a new result we have the proof of convergence of the procedure, and of existence of the diagonalisation matrix. We have also made explicit the issues with normalisation and hermiticity, and provided explicit solutions. We have shown by example the simplicity and effectiveness of the expansions in the \(\Lambda\) system. 

In the \(\Lambda\) system itself we have produced explicit formulae for the effective hamiltonian; in the text itself we have mentioned that only two types of terms ever appear ( \( \Omega^{\dag} \Omega \) and  \( \Omega^{\dag} \Omega \sigma_3\) for the non-hermitian effective hamiltonian). We expect that this kind of result will   generally be available in other systems, and we suggest that one of the advantages of the method put forward here is that it will facilitate the identification of such general structures. 

We have not made explicit the connection between our algebraic procedures and the resolvent techniques; we leave this for future work. We also postpone the application of these expansions to improve on the Rotating Wave Approximation, which is work in progress.

We have also not addressed the corresponding analysis for open quantum systems (for density matrices in closed open systems, see appendix \ref{app:mixed}); we hope our results will also be helpful there.
\acknowledgments
It is a pleasure to acknowledge support by  the Basque Government (IT-559-10), the Spanish Ministry of Science and Technology under Grant No. FPA2009-10612, the Spanish Consolider-Ingenio Program CPAN (Grant No. CSD2007-00042), and the UPV/EHU UFI 11/55. Thanks are also extended for fruitful conversations with G\'eza T\'oth, Manuel A. Valle Basagoiti, and Enrique Solano and his QUTIS group.

\appendix
\section{Bloch wave operator and Bloch's equation}\label{app:blochwave}
For completeness, we shall here make explicit the relation between our operator \( B \) and the Bloch wave operator in the literature. The Bloch wave operator is frequently denoted as \( \Omega \) or \( \Omega(E) \) (depending on whether one denotes the reduced or energy dependent wave operator), as, for example do Viennot  \cite{Viennot:2013fk} or Brandow \cite{RevModPhys.39.771}; Killingbeck and Jolicard \cite{0305-4470-36-20-201} denote it by \( W \), Bravyi et al. \cite{Bravyi20112793} by \( \mathcal{U} \). To avoid confusion with our use of the symbol, we shall make reference to the Bloch wave operator as \( \mathcal{B} \) and \( \mathcal{B}(E) \) respectively.

The Bloch wave operator fulfils the condition \( \mathcal{B}^2= \mathcal{B} \), and is defined such that \( P H \mathcal{B} P\psi=EP\psi\) for \( E \in \sigma(H)\) and \( P \) a projector. In fact, for \( \psi_0=P\psi_0\in \mathcal{H}_P \) in the eigenspace of the projector, it maps it to the full space, \( \mathcal{B}\psi_0\in \mathcal{H} \), such that the relevant eigenspace of \( H \) is invariant. That is,
\[ H \mathcal{B}= \mathcal{B} H \mathcal{B}\,,\]
which is called the Bloch equation. To see the equivalence  of this equation to eqn. (\ref{eq:bembedding}), observe that  \( \mathcal{B}^2= \mathcal{B} \) implies 
\[H \mathcal{B}= \mathcal{B} H \mathcal{B}\Leftrightarrow \left[ H, \mathcal{B}\right] \mathcal{B}=0\,.\]
Let \( P= \begin{pmatrix}
1&0\\0&0
\end{pmatrix}  \). Notice that the assignment
\[ \mathcal{B}= \begin{pmatrix}
1&0\\ B&0
\end{pmatrix}\]
satisfies 1) \( \mathcal{B}^2= \mathcal{B} \), 2) \( \mathcal{B}P= \mathcal{B} \), and inserting it into the Bloch equation in the right hand side form we obtain
\begin{multline}
 \left[ H, \mathcal{B}\right]\mathcal{B}= \left[ \begin{pmatrix}
 \omega& \Omega^{\dag}\\ \Omega& \Delta
 \end{pmatrix},\begin{pmatrix}
1&0\\ B&0
\end{pmatrix}\right]\begin{pmatrix}
1&0\\ B&0
\end{pmatrix}\\
=\begin{pmatrix}
\Omega^{\dag}B&- \Omega^{\dag}\\ \Delta B- B \omega + \Omega&- B \Omega^{\dag}
\end{pmatrix}\begin{pmatrix}
1&0\\ B&0
\end{pmatrix}\\
= \begin{pmatrix}
0&0\\ \Delta B - B \omega + \Omega - B \Omega^{\dag} B&0
\end{pmatrix}\,.
\end{multline}
\section{Gauge invariance}
A recurrent topic in the literature about extensions of the adiabatic elimination approximation is the dependence of the effective hamiltonian on the initial scheme. We shall now consider the change in the Bloch reduced wave operator \( B \) (also called decoupling operator or correlation operator) due to a class of changes of picture. Namely, consider the family of pictures given by unitary transformations that do not mix the slow and fast variables, i. e.  unitary transformations \( \tilde{\psi}=V\psi \) of the form
\[V(t)\psi= \begin{pmatrix}
v_{ \alpha}(t)&0\\0& v_{ \gamma}(t)
\end{pmatrix} \begin{pmatrix}
\alpha\\ \gamma
\end{pmatrix}\,,\]
with unitary blocks \( v_{ \alpha}(t) \) and \( v_{ \gamma}(t) \), 
\[ v_{ \alpha}(t)^{\dag}v_{ \alpha}(t)=1\,,\qquad v_{ \gamma}(t)^{\dag}v_{ \gamma}(t)=1\,.\]
The total hamiltonian in these pictures is
\[
\frac{1}{\hbar}\tilde{H}= \begin{pmatrix}
h_{ \alpha}+ v_{ \alpha} \omega v_{ \alpha} ^{\dag} &v_{ \alpha} \Omega^{\dag} v_{ \gamma} ^{\dag} \\
v_{ \gamma} \Omega v_{ \alpha} ^{\dag}& h_{ \gamma}+ v_{ \gamma} \Delta v_{ \gamma}^{\dag}
\end{pmatrix}\,,\]
where obviously 
\[h_{ \alpha}(t)= \left[i\partial_ t v_{ \alpha}(t)\right]  v_{ \alpha}^{\dag}(t)\,,\quad h_{ \gamma}(t)= \left[i\partial_ t v_{ \gamma}(t)\right]  v_{ \gamma}^{\dag}(t)\,.\]
 Let \( \tilde{B} \) be the corresponding Bloch reduced wave operator. After some algebra, the equation which determines it can be written as
 \begin{multline}\label{eq:gaugetranseq}
 \Delta \left( v_{ \gamma}^{\dag}\tilde{B}v_{ \alpha}\right) -  \left( v_{ \gamma}^{\dag}\tilde{B}v_{ \alpha}\right)  \omega + \Omega -  \left( v_{ \gamma}^{\dag}\tilde{B}v_{ \alpha}\right)  \Omega^{\dag}  \left( v_{ \gamma}^{\dag}\tilde{B}v_{ \alpha}\right) =\\
  \left( v_{ \gamma}^{\dag}\tilde{B}v_{ \alpha}\right)  \left( v_{ \alpha}^{\dag} h_{ \alpha} v_{ \alpha}\right)-  \left( v_{ \gamma}^{\dag} h_{ \gamma} v_{ \gamma}\right) \left( v_{ \gamma}^{\dag}\tilde{B}v_{ \alpha}\right) \,.
 \end{multline}
 If the second term were zero, then  \(  \left( v_{ \gamma}^{\dag}\tilde{B}v_{ \alpha}\right)  \) must be a solution for the original equation, so it equals the original \( B \). The new effective hamiltonian, under the same hypothesis, is simply
 \[\tilde{h}_{ \mathrm{eff}}= h_{ \alpha}+ v_{ \alpha} h_{ \mathrm{eff}} v_{ \alpha}^{\dag},,\]
 while the normalisation operator reads
 \[\tilde{S}_B= \sqrt{1+  v_{ \alpha} B B^{\dag} v_{ \alpha}^{\dag}}=  v_{ \alpha} S_B v_{ \alpha}^{\dag}\,.\]
 
 We do not have a full characterisation of the block diagonal unitaries for which the right hand side of eqn. (\ref{eq:gaugetranseq}) vanishes; we do know, however, of a case that has puzzled some researchers in the past (see the discussion in section 3 of  \cite{1751-8121-40-5-011}). Namely, when 
 \[ V(t)= \exp[-i \varphi(t)] \times \begin{pmatrix}
 1_{ \alpha}&0\\0&1_{ \gamma}
 \end{pmatrix}\,,\]
 since it follows that 
 \[h_{ \alpha} = \dot\varphi(t) P\quad \mathrm{and}\quad h_{ \gamma} = \dot\varphi(t) Q\,,\]
 from which \( \tilde{B}=B \) and
 \[\tilde{h}_{ \mathrm{eff}}=  \dot\varphi(t) 1_{ \alpha}+ h_{ \mathrm{eff}}\,,\]
 while \( \tilde{S}_B=S_B \).
 
 Summarizing, under a gauge transformation (time dependent phase transformation) the Bloch reduced wave operator is invariant, while the effective hamiltonian is covariant as expected (shifted by \( \dot\varphi(t) \)).
 
Similarly, for a constant change of basis that respects the structure \( \mathcal{H}= \mathcal{H}_{ P}\oplus \mathcal{H}_{ Q}  \), we have that \( B \), \( h_{ \mathrm{eff}} \) and \( S_B \) change rigidly.

\section{Another perturbative expansion}\label{app:seconperturb}
Consider now the situation in which the spectrum of \(\omega\) and the spectrum of \(\Delta\) are disjoint and such that the \(\Omega\) terms are perturbative; to be explicit, let us introduce an expansion parameter \(\epsilon\) by the substitution \(\Omega\to \epsilon \Omega\), and consider the reduced Bloch equation
\[ \Delta B- B \omega= - \epsilon \Omega + \epsilon B \Omega^{\dag} B\,.\]
\( B \) will be represented by a perturbation expansion of the form
\[B=\sum_{k=0}^\infty \epsilon^{2k+1}b_k\,.\]
Thus we rewrite Bloch's equation as the infinite set
\begin{eqnarray}
\Delta b_0- b_0 \omega&=& - \Omega\,,\nonumber\\
\Delta b_{k+1}- b_{k+1} \omega&=& \sum_{l=0}^k b_{k-l} \Omega^{\dag}b_l\,,\nonumber
\end{eqnarray}
for \( k\geq0 \). Each of the equations of this recurrence is a Sylvester equation, and (with some caveats in the infinite dimensional case) the solution is unique if \( \sigma( \Delta)\cap \sigma( \omega)=\emptyset \) \cite{Bhatia01011997}.
\begin{figure}[htbp]
\begin{center}
\includegraphics[width=2in]{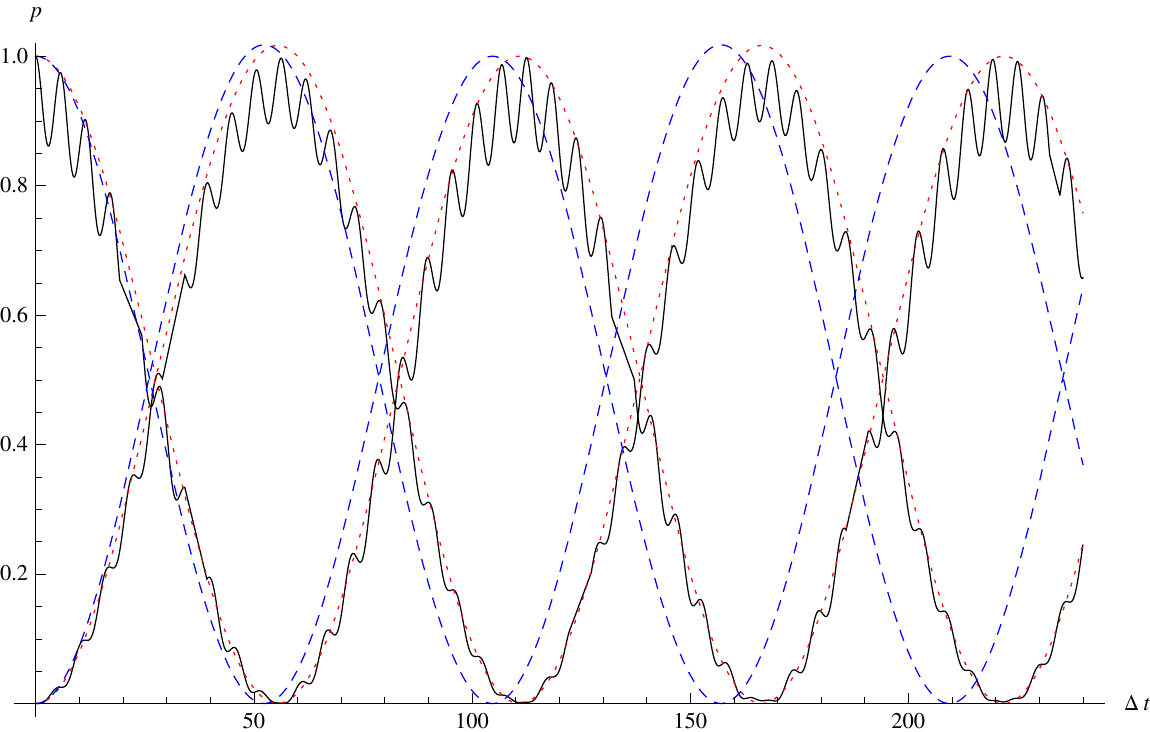}

\caption{Evolution of the population of the ground and excited states, with initial state \( (1,0,0)^T \), under a) the exact hamiltonian (continuous black line), b) \( h_{ \mathrm{eff}} \) obtained with \( B \to b_0\) (dashed blue line)  and c) \( h_{ \mathrm{eff}} \) with \( B=b_0+b_1+b_2+b_3 \) (dotted red line) The parameters are \( \delta= -0.0175 \Delta\), \( \Omega_a=0.4 \Delta \), \( \Omega_b=0.3 \Delta \), as before.}
\label{fig:sylvexpansion}
\end{center}
\end{figure}
In figure \ref{fig:sylvexpansion} we show that this expansion is also applicable to the numerical example used in the main text, with similar results.
This method would be preferable if \(\Delta\) were not so different from \( \pm \delta/2 \); if there is the wide separation between \(\omega\) and \(\Delta\) asked for in the text, this method provides a resummation of the perturbative adiabatic elimination expansion.

\section{No two cycle}\label{app:nocycles}
There is an appealing argument to discard the possibility of a two cycle for the \( T \) transformation. Assume that \( A_1 \) and  \( A_2 \) are related by a two-cycle, that is
\[T(A_1)=A_2\,,\quad T(A_2)=A_1\,.\] 
Then let 
\[A_+=A_1+A_2\,,\quad A_-=A_1-A_2\,,\]
which must obey
\begin{eqnarray}
\Delta A_+- A_+ \omega&=& - 2 \Omega + \frac{1}{2} A_+ \Omega^{\dag} A_++ \frac{1}{2}  A_- \Omega^{\dag} A_-\,,\nonumber\\
\left( \Delta+ \frac{1}{2}A_+ \Omega^{\dag}\right) A_-&=&- A_- \left( \omega + \frac{1}{2} \Omega^{\dag}A_+\right)\,.
\end{eqnarray}
The second expression has the form of a Sylvester equation. Given the conditions we have stated that guarantee the existence of a fixed point, the only solution for \( A_- \) in the second expression, taking \( A_+ \) as fixed, would be \( A_-=0 \), whence \( A_1=A_2 \), so there is no two cycle.

\section{Mixed states}\label{app:mixed}
The similarity transformation approach is very useful in that it has direct application to mixed states. As pointed out in the main text, given an operator \( A \)  acting on  the full Hilbert space we compute an effective operator acting on the slow, active, or model space as \( A_{ \mathrm{eff}} =PX^{-1}AXP\). Applying this to the density matrix we would obtain 
\[
\tilde{\rho}_{ \mathrm{eff}}= P X^{-1} \rho X P\,.
\]
This is not necessarily properly normalised, however. It makes sense to define, therefore,
\[ \rho_{ \mathrm{eff}}= \frac{1}{ \mathrm{Tr}\left[X P X^{-1} \rho(0) \right]}P X^{-1} \rho X P\,.\]
Notice that the normalization factor is the initial population of the low lying space in the unitary case.

If the evolution of the total system is hamiltonian and the decoupling equation (\ref{eq:decoupling}) holds, then the effective density matrix obeys the equation
\[
i\hbar \dot \rho_{ \mathrm{eff}}= \left[ H_{ \mathrm{eff}},\rho_{ \mathrm{eff}}\right]+ \frac{1}{ \mathrm{Tr}\left[X P X^{-1} \rho(0) \right]}P X^{-1}HX Q X^{-1} \rho X P\,.\]
As is only to be expected, the second term is zero if the decoupling equation holds and \( X \) is unitary.

An open question, which we hope to address in future work, is how this formalism ties in into the calculation of effective evolution for reduced open quantum systems performed by Reiter and S\o rensen \cite{Reiter:2012fk}. The proposal of extending the Schrieffer--Wolff approach to superoperators \cite{PhysRevA.86.012126} will undoubtedly prove relevant in this respect.


\end{document}